\documentclass[11pt]{article}

\begin{document}

\title{Exotic galilean symmetry and non-commutative mechanics
\\   
in mathematical \& in condensed matter physics
\footnote{Talk given at the Int. Conf. on Noncommutative Geometry and Quantum Physics. Kolkata, Jan. 2006.}}

\author{P.~A.~Horv\'athy 
 \\ 
 Laboratoire de Math\'ematiques et de Physique Th\'eorique 
 \\
 Universit\'e de Tours\\  Parc de Grandmont\\ 
 F-37200 TOURS (France)
\footnote{e-mail: horvathy@lmpt.univ-tours.fr.}}

\maketitle

\begin{abstract}
The ``exotic'' particle model with non-commuting position coordinates,
associated with the two-parameter central extension of the planar 
Galilei group, can be used to derive the ground
states of the Fractional Quantum Hall Effect. The relation to other
NC models is discussed. Anomalous coupling is presented.
Similar equations arise for a semiclassical Bloch electron, used to
explain the anomalous/spin/optical Hall effects.
\end{abstract}
\vspace{5mm}
\noindent
\texttt{hep-th/0602133}

\section{Introduction~: ``Exotic'' Galilean symmetry}
       
Central extensions first entered  
physics when Heisenberg realized that, in the quantum mechanics
of a massive particle, the position and
momentum operators did not commute. As a consequence, the group of 
space translations only acts up-to-phase
 on the quantum Hilbert space.  Expressed in  
more mathematical terms, it
 is not the [commutative] translation group itself, only its
[non-commutative] $1$-parameter central extension is
represented unitarily. 

Similarly, for a massive non-relativistic system
 Galilean boosts only act
up-to phase, so that it is its $1$-parameter central extension
 that acts unitarily.
True representations only arise for massless
particles.

Are there further extension parameters~? The question has been
asked and solved by Bargmann  \cite{Barg}~: in $d\geq3$ space dimensions,
the Galilei group only admits a $1$-parameter central
extension identified with physical mass, $m$. 
L\'evy-Leblond \cite{LL} has 
recognized, however that, 
owing to the commutativity of the planar rotation group $O(2)$, the Galilei group in the plane admits a second
``exotic''  extension, highlighted by
the non-commutativity of Galilean boost generators,
\begin{equation}
[K_1,K_2]=i\kappa, 
\label{exorel}
\end{equation}
where $\kappa$ is the new extension parameter.
 This fact has long been considered, however, a mere mathematical
curiosity, as
planar physics  has been viewed itself as a toy.
Around 1995 the situation started to change, though, with
the construction of physical models with such an ``exotic'' structure
 \cite{Grigore,LSZ}. 
These models have the strange feature that the Poisson bracket
of the planar coordinates does not vanish,
\begin{equation}
\{x_1,x_2\}=\frac{\kappa}{m^2}\equiv\theta.
\label{NCfreerel}
\end{equation}

Physical consequences, drawn in Ref. (\cite{DH})
are presented in Section \ref{exomech} below. 

Independently and around the same time, similar
structures were considered in condensed matter physics, 
namely for the  Bloch electron \cite{Niu},
where it has been argued that the semiclassical dynamics 
should involve a Berry curvature term which induces an
``anomalous'' velocity term of the same form as in the
``exotic'' model of Ref. (\cite{DH}).
Recent developments include the Anomalous \cite{AHE},
the Spin \cite{SpinHall}
and the Optical  \cite{Optical,OptiHall,SpinOptics} Hall effects.

Below we review the exotic point-particle model of Ref. (\cite{DH}),
followed by a brief outline of the semiclassical Bloch electron.
Let us note, in conclusion, that exotic Galilean symmetry can also been 
extended to non-commutative (Moyal) field theory
\cite{NCFT,ExoFT}.

\goodbreak

\section{``Exotic'' mechanics in the plane}\label{exomech}

Our present understanding of the Fractional
Quantum Hall Effect is
based on the motion of charged vortices in a magnetic field \cite{QHE}. 
Such vortices arise as exact solutions
in a field  theory of matter coupled to an abelian gauge field $A_\nu$,
whose dynamics is governed by the Chern-Simons term \cite{CSvortex,JP}.
Theory can be either relativistic or nonrelativistic.
For the latter, boosts commute, but
exotic Galilean symmetry
can be found in a Moyal-version of Chern-Simons
field-theory \cite{ExoFT}.

In Ref. (\cite{Grigore,DH}) Souriau's ``orbit method'' \cite{SSD}
was used to construct a classical system associated with
L\'evy-Leblond's
``exotic'' Galilean symmetry.  It has an ``exotic''
symplectic form and a free Hamiltonian,
\begin{eqnarray}
\Omega_0&=&dp_i\wedge dq^i+
\frac{1}{2}\theta\,\varepsilon_{ij}\,dp^i\wedge{}dp^j,
\label{symplectic}
\\[6pt]
H_0&=&\frac{\vec{p}^2}{2m}.
\label{freeham}
\end{eqnarray}
The associated free motions follow the usual straight
lines; the ``exotic'' structure only enters the conserved quantities,
namely the boost and the angular momentum,
\begin{equation}
    \begin{array}{ll}
    &j=\epsilon_{ij}x_ip_j+\frac{\theta}{2}\,{\vec{p}}\strut^2,
    \\[8pt]
    &K_{i}=mx_{i}-p_{i}t+m\theta\,\epsilon_{ij}p_{j}.
    \end{array}
\label{consquant}
\end{equation}
The ``exotic'' structure behaves hence
roughly as spin:
it contributes to some conserved quantities, but the
new terms are separately conserved.
The new structure does not 
seem to lead to any new physics.

The situation changes dramatically, though, if the particle 
is coupled to a gauge field.  Applying Souriau's prescription
\cite{SSD} yields indeed
\begin{equation}
\Omega=\Omega_0+eB\,dq_1\wedge dq_2,
\qquad
H=H_0+eV.
\end{equation}
The associated Poisson bracket 
then automatically satisfies the Jacobi identity.
The resulting equations of motion read
\begin{equation}
\begin{array}{rcl}\displaystyle
m^*\dot{x}_{i}
&=&
p_{i}-\displaystyle em\theta\,\varepsilon_{ij}E_{j},
\\[8pt]
\displaystyle
\dot{p}_{i}
&=&
eE_{i}+eB\,\varepsilon_{ij}\dot{x}_{j},
\end{array}
\label{DHeqmot}
\end{equation}
where $\theta=k/m^2$ is the non-commutative parameter and
we have introduced the \textit{effective mass} $m^*$
\begin{equation}
m^*=m(1-e\theta B).
\label{effmass}
\end{equation}

The novel features, crucial for physical applications, are two-fold~:
Firstly, the relation between velocity and momentum, (\ref{velrel}),
contains an  ``anomalous velocity'' term, so that  
$\dot{x}_i$ and $p_{i}$ are not in general parallel. 
The second one is the interplay between the exotic structure and the magnetic field, yielding the effective mass $m^*$ in (\ref{Lorentz}).

Equations (\ref{DHeqmot}) come from the Lagrangian
\begin{equation}
 \int\!({\bf p}-{\bf A}\,)\cdot d{\bf x}
-\frac{p^2}{2}\,dt
+
\frac{\theta}{2}\,{\bf p}\times d{\bf p}.
\label{exolag}
\end{equation}

When $m^*\neq0$,
(\ref{DHeqmot}) is also a Hamiltonian system, $\dot{\xi}=\{h,\xi^\alpha\}$,
with $\xi=(p_i,x^j)$ and Poisson brackets
\begin{equation}
\begin{array}{lll}
\{x_{1},x_{2}\}=
\displaystyle\frac{m}{m^*}\,\theta,
	\\[3mm]
	\{x_{i},p_{j}\}=\displaystyle\frac{m}{m^*}\,\delta_{ij},
	\\[3mm]
	\{p_{1},p_{2}\}=\displaystyle\frac{m}{m^*}\,eB.
\end{array}
\label{exocommrel}
\end{equation}

A remarkable property is that for \textit{vanishing effective mass}
$m^*=0$, i.e., when the magnetic field takes the
critical value
\begin{equation}
B=\frac{1}{e\theta},
\end{equation}
the system becomes singular. Then ``Faddeev-Jackiw'' (alias symplectic)
reduction yields an  essentially two-dimensional, simple system,
reminiscent of 
``Chern-Simons mechanics'' \cite{DJT}. The symplectic plane plays, simultaneously,
the role of both configuration and phase space. The only motions are those which follow a generalized Hall law;
quantization of the reduced system yields the ``Laughlin''
wave functions \cite{QHE}, which are in fact the ground states
in the Fractional Quantum Hall Effect (FQHE).

The relations (\ref{exocommrel}) diverge as $m^*\to0$,
but after reduction we get,cf.(\ref{NCfreerel}),
\begin{equation}
\{x_1,x_2\}=\frac{1}{eB}=\theta.
\label{NCrel}
\end{equation}

\section{Relation to another non-commutative mechanics}

The exotic relations (\ref{exocommrel}) are similar to 
those proposed in Ref. \cite{NaPo},
\begin{equation}
\begin{array}{lll}
\{x_{i},x_{j}\}
   =\theta\epsilon_{ij},
\\[6pt]
\{x_{i},p_{j}\}=\delta_{ij},
\\[6pt]
\{p_{1},p_{2}\}=eB.
\end{array}
\label{NaPocommrel}
\end{equation}
The eqns of motion $\dot{\xi}=\{\xi,H\}$, where $\xi=(p_i,x^j)$ and
$
H=\frac{p^2}{2m}+eV(x)
$
is the standard Hamiltonian, read
\begin{equation}
\begin{array}{ll}
mx_i'&=p_i-em\theta\epsilon_{ij}E_j, 
  \\[8pt]
 p_i'&=eB\epsilon_{ij}\displaystyle\frac{p_j}{m}+eE_i,
\end{array}
\label{NaPoeq}
\end{equation} 
where we noted ``time'' by $T$; $(\cdot)'=\frac{\ d}{dT}$.
Then a short calculation shows that
\begin{equation}
\Big\{x_i,\{p_1,p_2\}\Big\}_{cycl}=
e\theta\epsilon_{ij}{\partial}_jB,
\end{equation}
so that the Jacobi identity is only satisfied if $B=$ const.

How can this theory be extended to an arbitrary $B$~?

 $\bullet$ Let us first assume that $B=$ const. 
 s.t. $m^*\neq0$, and {\it let us redefine the time} \footnote{This was suggested to me by G. Marmo.}, as
\begin{eqnarray}
  T\to t=(1-e\theta B)T
  \qquad\Rightarrow\qquad
  \frac{\ d}{dT}=(1-e\theta B)\, \frac{\ d}{dt}\,.
  \label{rescale}
\end{eqnarray}

Then eqns. (\ref{NaPoeq}) carried into the exotic eqns
(\ref{DHeqmot}). It follows that the two theories are,
under these conditions, equivalent.

 $\bullet$ The crucial fact is that the
time redefinition actually 
extends the previous theory, since it carries
into the ``exotic model'', where the Jacobi identity
 holds for any, not necessarily constant $B$.

The model (\ref{NaPocommrel}) has another strange feature.
Let us indeed assume that the magnetic field is
radially symmetric, $B=B(r)$. One would then expect to have 
conserved angular momentum. For constant $B$, applying
Noether's theorem to an infinitesimal  rotation 
$
\delta \xi_i=\epsilon_{ij}\xi_j
$
yields indeed 
$
\delta \xi_i=-\{J^{NP},\xi_i\},
$ 
with
\begin{equation}
J^{NP}=\displaystyle\frac{1}{1-e\theta B}
\underbrace{\left(\vec{q}\times\vec{p}+
\frac{\theta}{2}\vec{p}{}^2+\frac{eB}{2}\vec{q}{^2}
\right)}_{j}\ .
\label{JNangmom}
\end{equation}
This differs from the standard expression in the pre-factor
\begin{eqnarray*}
\frac{1}{1-e\theta B}.
\end{eqnarray*}
For $B=B(r)\neq$ const., however, (\ref{JNangmom}) is
not in general conserved~:
$$
\frac{dJ^{NP}}{dT}=\frac{e\theta}{1-e\theta B}
\,\partial_iBx_i',
$$
while $j$ in (\ref{consquant})  is conserved.

\section{Physical origin of exotic structure}
 
A free  relativistic ``elementary'' particle in the plane corresponds to
a unitary representation of the planar Lorentz group $o(2,1)$. 
According to geometric quantization, these representations are associated
with the coadjoint orbits of the planar Lorentz group $SO(2,1)$, endowed with their canonical symplectic
structures. Following Souriau, these latters can in turn be viewed as
classical phase spaces.

For the planar Lorentz group, the procedure yields \cite{anyons}
 \begin{eqnarray}
 \Omega_0&=&dp_\alpha\wedge dx^\alpha+
\frac{s}{2}\epsilon^{\alpha\beta\gamma}
 \frac{p_\alpha dp_\beta\wedge dp_\gamma}
 {(p^2)^{3/2}},
 \\[8pt]
 H_0&=&\frac{1}{2m}\big(p^2-m^2c^2\big). 
 \end{eqnarray}

Then, as pointed out by Jackiw and Nair \cite{JaNa},
 the free exotic model  can be recovered considering a 
tricky non-relativistic  limit, namely
\begin{equation}
 s/c^2\to \kappa=m^2\theta.
 \end{equation}
Then $\Omega_0\Big|_{H_0=0}$   goes over into the
exotic symplectic form.
Intuitively, the exotic structure can be viewed as a ``non-relativistic
shadow'' of relativistic spin.
 
 At the level of the field equations, a similar procedure, applied to the
 infinite-component Majorana-type equation considered by
Jackiw \& Nair, or by Plyushchay \cite{anyons}
 yields a  first-order infinite-component ``L\'evy-Leblond type''  system
\cite{HP2}.

The exotic Galilei group can itself be derived from the planar Poincar\'e group
 by  ``Jackiw-Nair''  contraction \cite{JaNa}. One starts with the  planar Lorentz generators, 
\begin{equation}
 \{J^\alpha,J^\beta\}=\epsilon^{\alpha\beta\gamma}J_{\gamma}.
 \end{equation}
 For the classical system 
 \begin{equation}
 J_\mu=\epsilon_{\mu\nu\rho}x^\nu p^\rho+s\frac{p_\mu}{\sqrt{p^2}}.
 \end{equation}
Non-relativistic boost are the ``JN'' limits of
 \begin{equation}
 \frac{1}{c}\epsilon_{ij}J^j
 \to mx_i-p_it+m\theta\epsilon_{ij}p_j=K_i,
 \end{equation}
and the  exotic relation is recovered,
\begin{equation}
\{K_1,K_2\}=J_0/c^2\to 
\displaystyle\frac{s}{c^2}=\kappa.
 \end{equation}
The angular momentum is in turn 
 \begin{equation}
 J_0=\vec{x}\times\vec{p}+s+\frac{s}{m^2c^2}\vec{p}\strut{}^2
 \to \vec{x}\times\vec{p}+\frac{1}{2}\kappa\vec{p}{}^2=j.
 \end{equation}
whereas the divergent term $s=\kappa c^2$ has to be
removed by hand.
 
\section{Anyons in e.m. fields}

Chou, Nair, Polychronakos \cite{CNP} suggested to describe
an anyon in an electromagnetic field by the equations 
\begin{equation}
 \begin{array}{llll}
m\displaystyle\frac{dx^\alpha}{d\lambda}&=&
p^\alpha
\qquad
&\hbox{(velocity-momentum)}
\\[10pt]
\displaystyle\frac{dp^\alpha}{d\lambda}&=&\frac{e}{m}F^{\alpha\beta}p_\beta
 \qquad
 &\hbox{(Lorentz equation)}
\end{array}
\label{CNP}
\end{equation}
These equations are Hamiltonian, with symplectic form and
Hamilton's function
\begin{eqnarray}
 \Omega&=&\Omega_0+\frac{1}{2} eF_{\alpha\beta}dx^\alpha\wedge dx^\beta,
 \label{CNPsymp}
 \\[8pt]
 H&=&H_0+
 \frac{es}{2m\sqrt{p^2}}\epsilon_{\alpha\beta\gamma}
 F^{\alpha\beta\gamma}p^\gamma,
 \label{CNPHam}
 \end{eqnarray}
respectively. Let us observe that the second, non-minimal term
 in the Hamiltonian is dictated by the form of
the velocity relation in (\ref{CNP}). 
 
As proved by Chou et al. in Ref. (\cite{CNP}), their
model has gyromagnetic ratio $g=2$, which has long been 
believed by 
high-energy-physics theoreticians  \cite{CNP,anyong} to be
the ``correct'' $g$ value of anyons. Experimental evidence
 shows, however,  that in various condensed-matter
situations including the Fractional Quantum Hall Effect,
the measured value of $g$ is approximately zero\cite{g0}. 

Is it possible to construct an ``anomalous'' model with
$g\neq2$~? The answer is affirmative \cite{AnAn}.
  Planar spin has to satisfy the relation $S_{\alpha\beta}p^\beta=0$. 
The spin tensor has, therefore, the form
\begin{equation}
S_{\alpha\beta}=
\frac{s}{\sqrt{p^2}}\epsilon_{\alpha\beta\gamma}p^\gamma.
\end{equation}
Introducing the shorthand  $-F_{\alpha\beta}S^{\alpha\beta}=F\cdot S$,
the Hamiltonian  (\ref{CNPHam}) is presented as
\begin{eqnarray}
H^{CNP}=\frac{1}{2m}\Big(p^2-M^2c^2\Big)
\quad\hbox{where}\quad
 M^2=m^2+\frac{e}{c^2}F\cdot S.
 \label{CNPHambis}
\end{eqnarray}
Let us observe that the ``mass'' $M$ depends here on spin-field coupling.
Our clue for generalizing this model has been the formula put forward
by Duval more than three decades ago \cite{DThese}~: let us posit instead
of (\ref{CNPHambis}) the mass formula
\begin{equation}
 M^2=m^2+\displaystyle\frac{g}{2}\,\frac{e}{c^2}F\cdot S,
 \label{gM}
\end{equation}
where $g$ is an arbitrary real constant. Then {\it consistent
equations of motion} are obtained 
 for any $g$, namely
\begin{eqnarray}
D\frac{dx^\alpha}{d\tau}&=&
G\frac{p^\alpha}{M}+(g-2)\,\frac{es}{4M^2}\,
\epsilon^{\alpha\beta\gamma}F_{\beta\gamma},
\label{DHvel}
 \\[6pt]
\frac{dp^\alpha}{d\lambda}&=&\frac{e}{m}F^{\alpha\beta}p_\beta,
\label{DHLorentz}
\end{eqnarray}
where the coefficients denote the complicated, field-dependent expressions
\begin{equation}
D=1+\frac{eF\cdot S}{2M^2c^2},
\qquad
 G=1+\frac{g}{2}\,\frac{eF\cdot S}{2M^2c^2}.
\end{equation}

For the choice $g=2$ the generalized model plainly reduces to 
eqn. (\ref{CNP})  of Chou et al. in (\cite{CNP}).

We can now consider the ``Jackiw-Nair'' non-relativistic limit of the above
relativistic model. This provides us, for  any $g$, with the 
Lorentz eqn. (\ref{Lorentz}), supplemented with
\begin{equation}
(M_gD)\dot{x}_i=
Gp_i-\big(1-\frac{g}{2}\big)
eM_g\theta\epsilon_{ij}E_j,
\end{equation}
where
\begin{eqnarray*}M_g=m(\sqrt{1-g\theta eB}),
\quad
D=\big(1-(g+1)\theta eB\big),
\;
G=\big(1-(3g/2))\theta eB\big).
\end{eqnarray*}

$\bullet$ It is a most important fact that,
for {\it any} $g\neq2$, the only consistent motions follow a
generalized Hall law, whenever the field takes {\it either}
of the critical values
\begin{equation}
B=\frac{1}{1+g}\,\frac{1}{e\theta}
\qquad\hbox{or}\qquad
\frac{2}{3g}\,\frac{1}{e\theta}.
\end{equation} 
One can indeed show that, for any $g\neq2$, the models can
be transformed into each other by a suitable redefinition.
For $g=0$ the equations become identically satisfied.
See \cite{AnAn} for details.

$\bullet$ In particular, for $g=0$ the minimal exotic  model of 
Ref. (\cite{DH})
is obtained. The latter is, hence, {\it not} the  NR limit of the model of
\cite{CNP} (\ref{CNP}) which has, as said, $g=2$.
The experimental evidence in \cite{g0} is, hence, a strong argument
in favour of the minimal model of Ref. (\cite{DH}).
 
$\bullet$ $g=2$ is in fact the only case, when 
the velocity \& the momentum are parallel. This is, however,
{\it not} required by any first principle, as advocated
a long time ago~: a 
perfectly consistent model is obtained for any $g$ 
\cite{DThese,Dixon,Souriau74}. Having non-parallel velocity and momentum
seems to be unusual in high-energy physics; it is, however, a well accepted
and even crucial requirement in condensed matter physics, as explained 
in the next Section.

\section{The semiclassical Bloch electron}

Around the same time and with no relation to the
above developments, 
a very similar theory has arisen in solid state physics \cite{Niu}. 
Applying a Berry-phase argument to a Bloch electron in a lattice,
 a semiclassical model can be derived \cite{Niu};
the equations of motion in the $n{}^{th}$ band read 
\begin{eqnarray}
\dot{\bf r}&=\displaystyle\frac{\partial\epsilon_n({\bf p})}
{\partial{\bf p}}-\dot{\bf p}\times{\bf \Theta}({\bf p}),
\label{velrel}
\\[6pt]
\dot{{\bf p}}&=-e{\bf E}-e\dot{{\bf r}}\times{\bf B}({\bf r}),
\label{Lorentz}
\end{eqnarray}
where ${\bf r}=(x^i)$ and ${\bf p}=(p_j)$ denote the electron's three-dimensional
intracell position and quasimomentum,
respectively, $\epsilon_n({\bf p})$ is the band energy. The purely momentum-dependent 
${\bf \Theta}=(\Theta_i)$ is the Berry curvature of the electronic Bloch states,
$\Theta_i({\bf p})=\epsilon_{ijl}\partial_{{\bf p}_j}a_l({\bf p})$, 
where $a_i$ is the Berry connection. 
A non-trivial Berry connection requires broken time-reversal symmetry,
as it happens, e. g., in GaAS heterostructures \cite{Niu}.

Recent applications of the model include the Anomalous \cite{AHE} 
and the Spin \cite{SpinHall}  Hall Effects.
All these developments are based on the anomalous velocity term
\begin{equation}
\dot{\bf p}\times{\bf \Theta}({\bf p})
\label{anovel}
\end{equation}
that corresponds to  the anomalous current
 advocated by Karplus and Luttinger
 as long as fifty years ago \cite{AHE} to explain
 the Anomalous Hall Effect, observed  in some ferromagnetic
 matter in the absence of a magnetic field.
 Now, as confirmed experimentally by Fang et al.
in Ref. (\cite{AHE}), in the Anomalous Hall Effect 
the Berry curvature can take the form
of a {\it monopole in ${\bf p}$-space},
\begin{equation}
{\bf\Theta}=g\frac{\bf p}{p^3},
\label{pmonop}
\end{equation}
which is indeed the only possibility consistent with spherical
symmetry \cite{BeMo}. 
For ${\bf B}=0$ we have $\dot{\bf p}=-e{\bf E}$.
Then, taking the parabolic case $\epsilon_n({\bf p})$
for simplicity, the velocity relation (\ref{velrel}) becomes
\begin{eqnarray}
\dot{\bf r}={\bf p}
+\frac{eg}{p^3}{\bf E}\times{\bf p}.
\label{AHEvelrel}
\end{eqnarray}
The anomalous term shifts the velocity and deviates, hence, the particle's trajectory perpendicularly
to the electric field -- just like in the ordinary
Hall effect.
      
Eqns. (\ref{velrel}-\ref{Lorentz})  derive from the Lagrangian
\begin{equation}
     L^{Bloch}=\big(p_{i}-eA_{i}({\bf r},t)\big)\dot{x}^{i}-
     \big(\epsilon_n({\bf p})+eV({\bf r},t)\big)
     +a^{i}({\bf p})\dot{p}_{i},
     \label{blochlag}
\end{equation}
and are also consistent with the Hamiltonian structure
\cite{DHHMS}
\begin{eqnarray}
\{x^i,x^j\}^{Bloch}&=\displaystyle\frac{\epsilon^{ijk}
\Theta_k}{1+e{\bf B}\cdot{{\bf \Theta}}},
\label{xx}
\\[6pt]
\{x^i,p_j\}^{Bloch}&=\displaystyle\frac{\delta^{i}_{\ j}
+eB^i\Theta_j}{1+e{\bf B}\cdot{{\bf \Theta}}},
\label{kk}
\\[6pt]
\{p_i,p_j\}^{Bloch}&=-\displaystyle\frac{\epsilon_{ijk}eB^k}{1+e{\bf B}\cdot{{\bf \Theta}}}
\label{xk}
\end{eqnarray}
and Hamiltonian $h=\epsilon_n+eV$ \cite{BeMo}. 

Restricted to the plane, these equations reduce
to the exotic equations (\ref{DHeqmot}), provided 
$\Theta_i=\theta\delta_{i3}$. For $\epsilon_n({\bf p})={\bf p}^2/2m$
and chosing $A_i=-({\theta}/{2}) \epsilon_{ij}p_j$, 
the semiclassical Bloch Lagrangian
(\ref{blochlag}) becomes the ``exotic'' expression (\ref{exolag}).

The exotic galilean symmetry is lost if $\theta$ is not constant,
though.

A similar pattern arises in optics \cite{Optical,OptiHall,SpinOptics}~:
to first order in the gradient of the refractive index $n$, spinning
light is approximately described by the equations
\begin{eqnarray}
\dot{\bf r}\approx{\bf p}-\frac{s}{\omega}\,{\rm grad }(\frac{1}{n})\times{\bf p},
\qquad
\dot{\bf p}\approx 
-n^3\omega^2{\rm grad }(\frac{1}{n}).
\label{LinDuv}
\end{eqnarray}
where $s$ denotes the photon's spin.
Here we recognize once again an anomalous velocity
relation of the type (\ref{velrel}). The new term
makes the light's trajectory deviate from that predicted
in ordinary geometrical optics, giving rise to an
``optical Magnus effect'' \cite{Optical}. A manifestation of this
is the displacement of the light ray perpendicularly to the
plane of incidence at the interface of two media with different
refraction index~: 
this is the ``Optical Hall Effect \cite{OptiHall,SpinOptics}.

\section*{Acknowledgments}
 This review is based on joint research with
C. Duval, Z. Horv\'ath, L. Martina, M. Plyushchay and
P. Stichel, to whom I express my indebtedness.


\end{document}